\documentclass[aps,prl,twocolumn,groupedaddress]{revtex4}

\usepackage[latin1]{inputenc}
\usepackage{amsmath}
\usepackage{verbatim}
\usepackage{amssymb}
\usepackage{bbm}
\usepackage{graphicx}

\newcommand{\approptoinn}[2]{\mathrel{\vcenter{
  \offinterlineskip\halign{\hfil$##$\cr
    #1\propto\cr\noalign{\kern2pt}#1\sim\cr\noalign{\kern-2pt}}}}}

\begin{document}


\title{Hopf-type neurons increase input-sensitivity by forming forcing-coupled ensembles }


\author{Florian Gomez, Tom Lorimer, and Ruedi Stoop}
\affiliation{Institute of Neuroinformatics and Institute of Computational Science, University and ETH Zurich, 8057 Zurich, Switzerland}


\date{\today}

\begin{abstract}
Astounding properties of biological sensors can often be mapped onto a dynamical system in the vicinity a bifurcation. For mammalian hearing, a Hopf bifurcation description has been shown to work across a whole range of scales, from individual hair bundles to whole regions of the cochlea. We reveal here the origin of this scale-invariance, from a general level, applicable to all neuronal dynamics in the vicinity of a Hopf  bifurcation (embracing, e.g., Hodgkin-Huxley equations). When coupled by natural 'force-coupling', ensembles of Hopf oscillators below bifurcation threshold exhibit a collective Hopf bifurcation. This collective Hopf bifurcation occurs {\em  substantially below} where the average of the individual oscillators would bifurcate, with a frequency profile that is sharpened if compared to the individual oscillators. \end{abstract}


\maketitle

Biological sensors often deal with inputs across many orders of magnitude (expressed by a logarithmic stimulus scale, e.g., dB or pH scales). They usually have the ability to strongly amplify weak inputs and to compress higher input levels. Prominent manifestations are the 'compressive nonlinearity' of the hearing system \cite{Hudspeth2008,Ashmore2010} and the nonlinear 'I/f curves' of neuronal response. Such properties emerge naturally from dynamical systems in the vicinity of, but mostly below, a bifurcation point (a bifurcation is a mathematical term describing a structural change of a solution of a dynamical system). This was already known thirty years ago, then termed 'small signal amplification' \cite{WiesenfeldMcNamara1985,WiesenfeldMcNamara1986,Derighetti1985}. Its important role as the working principle of biological sensors was recognized much later \cite{Eguiluz2000,Camalet2000}, and it was shown that this mechanism plays a decisive role in insect hearing \cite{Stoop2006,Lorimer2015a}. Regarding mammalian hearing \cite{KernStoop2003,Magnasco2003,Hudspeth2008}, this principle still is fighting its way against classical engineering hearing solutions.  Also for networks of neurons, which naturally display bifurcation behavior, corresponding consequences have largely remained unexplored. Here, we demonstrate novel phenomena of collective behavior of potential biological significance that emerge from this phenomenon.   
\begin{figure}[h!!!!!!!!!!!!!!!!!!!!!!!!!!!!!!!!!!!!!!]
\begin{center}
\includegraphics[width=1.0 \linewidth]{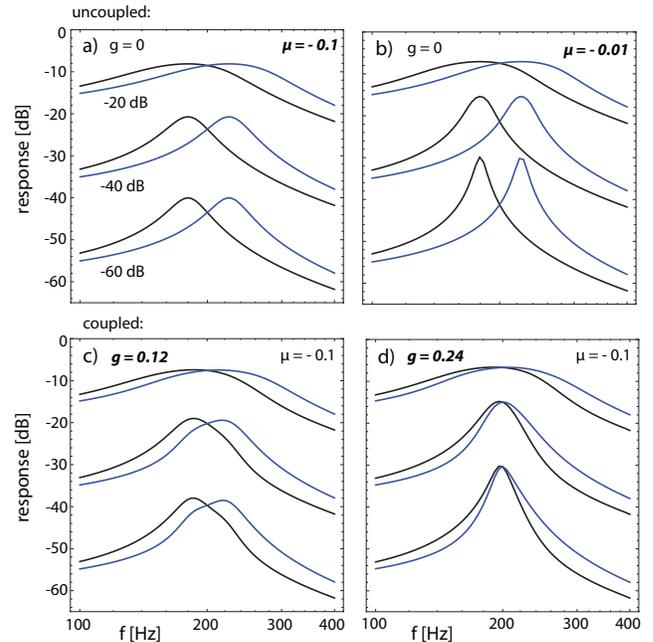}
\end{center}
\caption{Response of oscillators 1 and 2 (black/blue) to forcing of frequencies $f=\omega_{1,2}/2\pi=180, \,225$ Hz,  respectively, at three amplitudes. Uncoupled oscillators ($g=0$) : a), b). Coupled oscillators : c) $g=0.12$, d) $g=0.24$.} 
\label{fig1}
\end{figure}

Simulations of neuronal networks have to date focused on synchronization of pulse coupled phase oscillators \cite{Restrepo2006,SendinaNadal2008,Girnyk2012} or integrate and fire neurons \cite{LuccioliPoliti2010,Olmi2010,OKeeffe2015} that do not incorporate a Hopf bifurcation regime. In nature, different types of bifurcations occur that mathematics classifies according to the eigenvalues that the system's linearization has when the behavioral change occurs, i.e., at the bifurcation point. The Hopf bifurcation is among the most fundamental bifurcations in physics and biology \cite{HH,Eguiluz2000,KernStoop2003,Magnasco2003,Hudspeth2008}. In addition to the small signal amplifier properties, it has a sharp tuning regarding stimulation frequency (a feature shared by some other, more specific, bifurcations). 'Hopf systems' (for short) are at rest below the bifurcation point. When pushed by the 'right' stimulus, they amplify it, the stronger the closer they are to the bifurcation point (Fig. 1). If, due to a suitable external parameter change, they cross the bifurcation point, they start to oscillate (or 'spike') in a self-sustained manner, at a system-characteristic frequency. In the following, we will switch between the terms 'system', 'neuron', or 'oscillator', depending on the focus we want to convey.

In the cochlea, single outer hair cells as well as mesoscopic cochlea elements of hundreds of hair cells  \cite{KernStoop2003,Martignoli2007} are reliably described by a Hopf system. In this letter, we elucidate how a single element and ensembles composed of these elements adhere to the same physical description, and what this entrains further.  We will show that upon increased coupling, an ensemble of subthreshold Hopf oscillators merges into a collective state of synchronized behavior, showing ensemble coherent small-signal amplification behavior. Upon a further increase of the coupling, the ensemble may undergo a Hopf bifurcation, where the oscillators lock, and spontaneously oscillate in a synchronized manner. 
In both regimes, the behavior of the ensemble is indistinguishable to that of a single Hopf oscillator. Most astonishingly, however, is that this happens with a significantly sharpened frequency response profile and a much increased input sensitivity, if compared to the contributing systems (Fig. 1, a) vs. d)). Such behavior is not only fundamentally different from the bulk of studies focusing on coupled oscillators (e.g. \cite{Aronson1990,Strogatz2000, ErmentroutKleinfeld2001}), it also contradicts most every-day experience: Ensembles of cars on the street, e.g., do not exhibit such a sharpened reaction profile. 


To demonstrate this behavior, we start from the frequency-rescaled Hopf equation 
\begin{equation}
\dot{z} = (\mu + i) \,\omega_{ch}z -\omega_{ch}|z|^2 z +\omega_{ch} f(t),
\label{hopf}
\end{equation}
where $z(t)$ is the complex-valued variable, $\mu$ is the distance to the bifurcation point, $\omega_{ch}$ is the characteristic (angular) frequency of the oscillator, and $f(t)$ is a complex-valued external forcing input \cite{Martignoli2007}. In this formulation, a Hopf bifurcation occurs when the parameter $\mu$ changes from negative to positive values: For $\mu>0$, the system starts to spontaneously oscillate at a frequency $\omega_{ch}$. Below bifurcation (i.e. for $\mu<0$), the system is  quiet, but when stimulated by a signal $f(t,\omega)$, the stimulation is amplified, more strongly the closer $\omega$ is to $\omega_{ch}$ and the closer the system is to bifurcation. 
The simplest biological coupling of two such oscillators is by 'forcing-coupling'  
\begin{align}
\dot{z_1}&= \omega_1\left((\mu_1 + i) z_1 -|z_1|^2 z_1 + f_1(t)+ \frac{g_{21}}{2} z_2\right), \\
\dot{z_2}&= \omega_2\left((\mu_2 + i) z_2 -|z_2|^2 z_2 + f_2(t)+ \frac{g_{12}}{2} z_1\right), \nonumber
\label{hopf2}
\end{align}
where $\omega_{1,2}$ are the characteristic frequencies of the oscillators and $g_{ij}$ denotes the coupling from oscillator $i$ to oscillator $j$. This situation can easily be generalized; the factor of $1/2$ is introduced to facilitate the generalization to $N$ oscillators later on. 
Fixing the characteristic frequencies and keeping both oscillators at $\mu_1=\mu_2=\mu<0$, we first analyze the system's behavior for variable symmetric coupling strength $g=g_{12}=g_{21}$. To this end, we stimulate both oscillators with the single-frequency input $f(t)=f_1(t)=f_2(t)=f e^{i \omega t}$ and measure the responses for different values of $g$ and $\omega$. For $\omega_1/2\pi=180$ Hz, $\omega_2/2\pi=225$ Hz and $\mu=-0.1$, the result is displayed in Fig. \ref{fig1}a), c), d).
The evident effect of forcing-coupling is a transition from two individual, rather broad, response profiles to two almost overlapping sharply peaked profiles at $g=0.24$. The pronounced small-signal amplification characteristics around the center frequency $f_c=200$ Hz is the manifestation of a Hopf bifurcation of the {\em forcing-coupled} system. Keeping the $\mu$-parameters fixed (here at $\mu=-0.1$),  the coupling $g$ now takes the role of the bifurcation parameter: At a 'critical coupling' $g_c$, the origin $z_1=z_2=0$ loses stability; beyond $g_c$, the ensemble of the two oscillators starts to oscillate spontaneously at common frequency $f_c$. Indeed, close to bifurcation (here: $g_c\approx 0.3$), the response curves are comparable to that of an individual oscillator at $\mu \simeq -0.01$ (see Fig. \ref{fig1}b). The forcing-coupled system shows, however,  a sharper frequency tuning, focused around a common center frequency. 

To extract the critical coupling $g_c$ for different values of $\omega_{1,2}$ and $\mu$, we rewrite the system as a four-dimensional real-valued system and find for the Jacobian at the origin
\begin{equation}
J=\begin{pmatrix}
	\mu\, \omega_1 & -\omega_1 & \frac{g}{2}\omega_1 & 0  \\
	\omega_1 & \mu\, \omega_1 & 0 &  \frac{g}{2} \omega_1 \\
	 \frac{g}{2} \omega_2 & 0 & \mu\, \omega_2 & -\omega_2 \\
	0 & \frac{g}{2}\omega_2 & \omega_2 & \mu\, \omega_2
\end{pmatrix}.
\label{Jacobi}
\end{equation}
\begin{figure}[h!!!!!!!!!!!!!!!!!!!!!!!!!!!!]
\begin{center}
\includegraphics[width=0.8\linewidth]{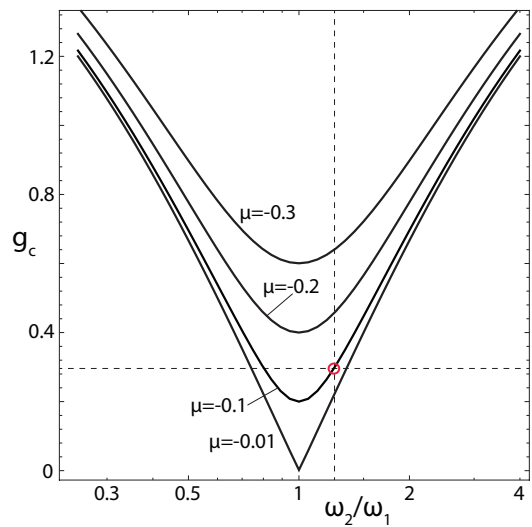}
\end{center}
\caption{Critical value $g_c$ for symmetric coupling (Eq. 5) $g=(g_{12}=g_{21})$, as a function of $\omega_2/\omega_1$ and four different values of $\mu$, for $\omega_1/2\pi=180$ Hz. The lines single out the regions above which the coupled system starts to oscillate, based on elements that are individually below bifurcation threshold. Red circle: Location of $g_c$ for $\omega_{1,2}/2\pi=180, \,225$ Hz.} 
\label{fig2}
\end{figure}
The stability of the origin is determined by the real part $\Re$ of the eigenvalues of the Jacobian. At the Hopf bifurcation, two eigenvalues simultaneously cross the imaginary axis, which can be used to determine the critical coupling value $g_c$. After some simplifications, we arrive at the implicit equation
\begin{equation}
(\omega_1+\omega_2) \mu =-\Re \left( \sqrt{(\mu^2-1+2\mu i) (\omega_1^2-\omega_2^2)+g_c^2\omega_1\omega_2} \right).
\label{criticalcoupling}
\end{equation}
Solving this equation numerically for $g_c$ yields a function $g_c(\omega_2/\omega_1,\mu)$. The obtained results (Fig. \ref{fig2}), express that coupling strongly enhances the emergence of a (collective) oscillation.
If the two oscillators have identical characteristic frequencies $\omega_1=\omega_2$, we obtain from Eq. (\ref{criticalcoupling}) $g_c=-2\mu$, which determines the location of the minima of the curves. Pushing either oscillator frequency into one direction increases $g_c$. Similarly, pushing $\mu$ further away from the bifurcation point $\mu=0$ shifts $g_c$ to higher values. For all couplings $g$, we may cross the bifurcation point by changing $\mu$. In this sense, $\mu$ does not fully abandon its role as a bifurcation parameter. However, the critical value is no longer $\mu_c=0$ of the uncoupled oscillators, but shifts to more negative values, expressing that the coupled system is more excitable than its components.

\begin{figure}[h!!!!!!!!!!!!!!!!!!!!!!!!!!!!!!!!!!!!!]
\begin{center}
\includegraphics[width=0.8\linewidth]{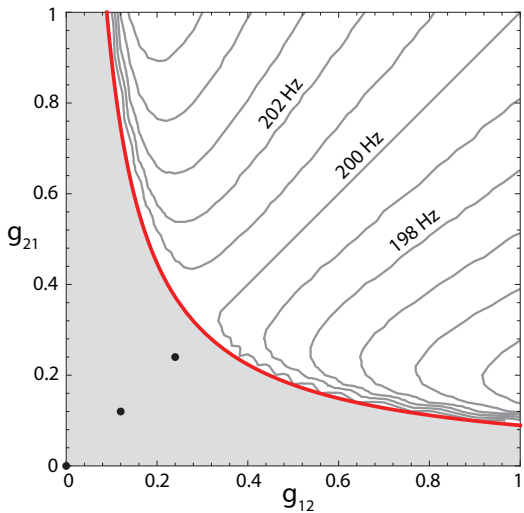}
\end{center}
\caption{Critical coupling $g_c$ (red) and frequency of oscillation (contours) above bifurcation for asymmetrical coupling  ($\omega_{1,2}/2\pi=180, 225$ Hz, $\mu_{1,2}=-0.1$). Gray-shaded area: small-signal amplification regime. Black dots: location of the systems of Fig. \ref{fig1}a), c) d).} 
\label{fig3}
\end{figure}

For asymmetrical coupling, $g_c^2$ in Eq. (\ref{criticalcoupling}) has to be replaced by $g_{12}\, g_{21}$. For, e.g., $g_{12}$ fixed, we easily find the critical value of $g_{21}$  (cf. Fig. \ref{fig3}, where in the $g_{12}$-$g_{21}$-space the critical line, together with the oscillation frequency after the bifurcation, is displayed). For symmetric coupling, the center frequency $f_c$ is also the frequency of the oscillation after bifurcation; it also coincides with the stimulation frequency to which the individual uncoupled oscillators would respond with equal strength (c.f. Fig. \ref{fig1}, where $f_c=200$ Hz, see below). 
By writing the uncoupled oscillators locked to an external signal of frequency $\omega$ as $a_j e^{i \omega t}$ with complex amplitudes $a_j$, $j=1,2$, we arrive at an explicit expression for $f_c$ of the coupled system. Inserting this into the Hopf equation (\ref{hopf}), we obtain
\begin{align}
a_j(i(\omega-\omega_j)-\omega_j \mu+\omega_j |a_j|^2)&=\omega_j f.
\label{eqhu1}
\end{align}
Multiplying both sides with the complex conjugates, we obtain for the responses $x_{1,2}:=|a_{1,2}|^2$ the cubic equations 
\begin{align}
x_1^3-2\mu x_1^2+\left(\mu+\frac{(\omega-\omega_1)^2}{\omega_1^2}\right) x_1&=|f|^2,\\
x_2^3-2\mu x_2^2+\left(\mu+\frac{(\omega-\omega_2)^2}{\omega_2^2}\right) x_2&=|f|^2.\nonumber
\label{eqhu2}
\end{align}
Requiring $x_1=x_2$ leads to (without having to explicitly solve these equations \cite{Eguiluz2000,KernStoop2003})
\begin{equation}
\frac{(\omega-\omega_1)^2}{\omega_1^2}=\frac{(\omega-\omega_2)^2}{\omega_2^2},
\label{eqhu3}
\end{equation}
from which the center frequency $f_c$ emerges as
\begin{equation}
\omega_c=2\pi f_c=\frac{2\omega_1\omega_2}{\omega_1+\omega_2}.
\label{eqhu4}
\end{equation}
For the example provided above ($\omega_1/2\pi=180$ Hz and $\omega_2/2\pi=225$ Hz, $\mu=-0.1$), this indeed yields $f_c=200$.  The forcing-coupled system keeps oscillating at $f_c$ as long as $g_{12}=g_{21}$. For  asymmetrical coupling, the oscillation frequency shifts towards the dominant oscillator's characteristic frequency (c.f. Fig. \ref{fig3}).

\begin{figure}[h!!!!!!!!!!!!!!!!!!!!!!!!!]
\begin{center}
\includegraphics[width=1\linewidth]{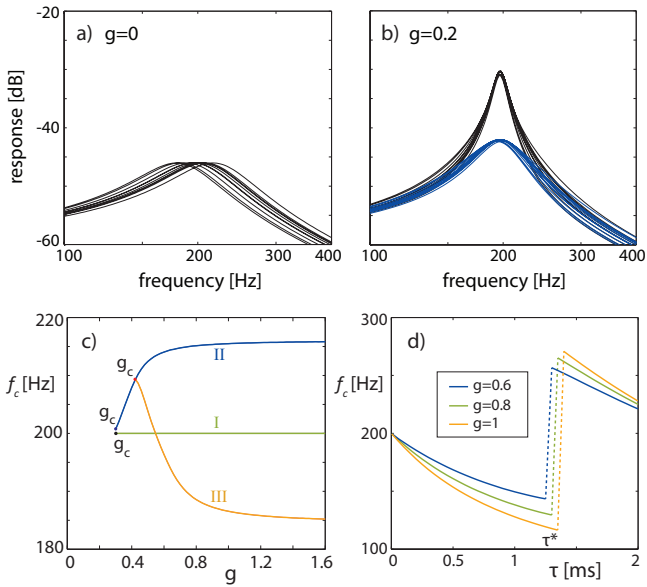}
\end{center}
\caption{Response of $N=10$ oscillators, characteristic frequencies distributed around 200 Hz. a) Uncoupled, $\mu=-0.2$, b) forcing-coupled (black: $\mu=-0.2$, gray: $\mu=-0.3$), exhibiting  a coherent and sharply tuned response around $f_c\approx 200$ Hz. c) $f_c$ as a function of the coupling $g$ beyond $g_c$ ($\mu=-0.1$ for all oscillators). I: two oscillators (180/225 Hz), II: three oscillators (180/200/300 Hz), III: five oscillators (120/160/200/240/300 Hz). d) $f_c$ for $\tau$-delayed coupling (two oscillators at 180/225 Hz, $\mu_{1,2}=-0.1$, $g=0.8$).} 
\label{fig4}
\end{figure}

For investigating how the ensemble Hopf bifurcation point depends on the ensemble size $N$, we keep the all-to-all forcing-coupling scheme (which implies replacing $g/2$ by $g/N$). Our finding is that, even for large $N$, the ensemble Hopf-bifurcation persists. In Fig. \ref{fig4}a),b),  the emergence of a coherent response profile, with response strengthening and sharpening upon the increase of the coupling, is exemplified for an ensemble of $N=10$ oscillators, with frequencies distributed around 200 Hz. 
Beyond bifurcation, the center frequency $f_c$ becomes a more complicated function of the coupling. While for two oscillators $f_c$ was constant (I), for more oscillators, $f_c(g)$ can either increase (II) or decrease (III) with $g$, until saturation dominates (Fig. \ref{fig4}c). This behavior resembles that of ensembles of synchronized {\em realistic} neurons, where the common synchronization frequency is a function of the coupling strength (very much in contrast to coupled Kuramoto phase oscillators, where an {\em invariable} synchronization frequency is observed \cite{Ferrari2015}). In our paradigm, $f_c$ depends on the distribution of the oscillator frequencies $\omega_i$. Only for perfectly symmetrical situations (e.g. 180/200/225 Hz or 100/100/200/200 Hz), we have $f_c=const$. A bias in the coupling changes the relative dominance among the oscillators and introduces a change in $f_c$. 
\begin{figure}[htb]
\begin{center}
\includegraphics[width=1\linewidth]{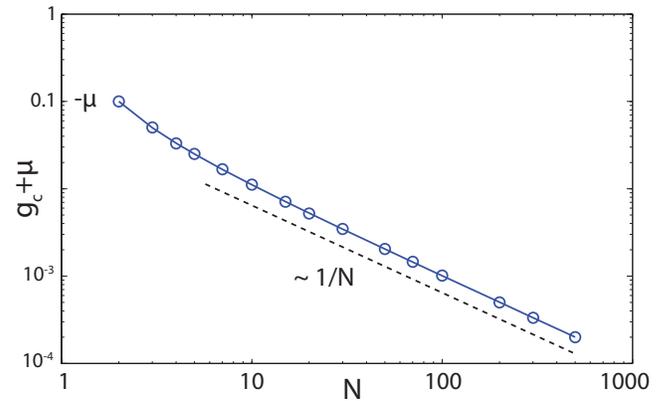}
\end{center}
\caption{$g_c+\mu$ as a function of the oscillator number $N$. 
While the presented result is for $\omega_{ch, \,i}=\omega_{ch}$ and $\mu_i=\mu=-0.1$, 
we observe the same behavior (with a line-shift) for reasonable variations of  $\omega_{ch, \,i}$ and $\mu_i$.
The value of $\mu$ is reflected in the first data point obtained for $N=2$, which implies $g_c=-2\mu$ (c.f. Eq. (5) and Fig. 2).}
\label{fig5}
\end{figure}

Our numerical results demonstrate that ensembles of $N$ forcing-coupled Hopf oscillators reproduce the small-signal amplification characteristics of single Hopf oscillators at augmented excitability, following a simple scaling relation of exponent $-1$ (Fig. \ref{fig5}). Without requiring a precise building principle, forcing-coupling (Eqs. (2) ) drives the ensemble towards a common characteristic frequency $f_c$ and establishes a coherent response profile that is largely independent of the individual oscillators' frequency distribution. We believe that such couplings are much more biologically relevant than the commonly used diffusive coupling paradigm that has its origins in an abstract equilibrium-physics point of view. Diffusive coupling (for two oscillators a coupling of the form $\sim \frac{g_{21}}{2}(z_2-z_1)$) would show a gradual convergence towards a center frequency $f_c$ as well, regarding excitability, it would, however, {\em stabilize} the system, instead of pushing it closer towards bifurcation. To arrive at collective self-oscillations, the individual bifurcation values would then have to be augmented to significantly positive values (typically $\mu>0.1$). 
Our forcing-coupling results are consistent regarding the inclusion of a natural delay of action $\tau$, where, for two oscillators, the couplings in Eqs. (2) would be replaced by $\frac{g_{21}}{2}\, z_2(t-\tau)$ and $\frac{g_{12}}{2}\,z_1(t-\tau)$, respectively. For $N$ oscillators, we considered that each oscillator obtains from all other oscillators the same time-delayed signal. This is a simple setting, in particular from the cochlear perspective, but is sufficient to assess the general effect of a signal delay. For small delays ($\tau<<1$ ms, which we expect to hold for neural information processing, the system properties persevere, in particular the small-signal amplification characteristics (Fig. \ref{fig4}d). Only for rather long delays of the order of 1ms,  $f_c$ gradually decreases. This behavior parallels observations known from systems of coupled oscillators \cite{Niebur1991}. At an even enlarged delay $\tau^*(g)$, a discontinuous jump of $f_c$ occurs, which repeats upon further increased delays, a phenomenon that resembles the change of locking observed in driven oscillators. The critical coupling value $g_c$ depends only mildly on $\tau$, in contrast to the behavior shown by $f_c$.

Hopf oscillators are prototypes of biological excitability \cite{HH, excitable1,excitable2,excitable3}. It is only upon an appropriate input from other nodes, often strongly dependent on external sources, that these elements are activated, a property that is shared by most biological neurons. Moreover, our setting reproduces observed biological synchronization frequency behavior. While the presently prominent modeling of neuronal ensembles by diffusively coupled Kuramoto phase-oscillators \cite{CuminUnsworth2007,Breakspear2010,HongStrogatz2011}, misses these important aspects of the biological behavior, models based on flexible bifurcation parameters as used here provide a more general approach, having the potential for a more thorough understanding of processing and propagation of information in biological networks.
By relaxing all-to-all to local coupling, we observed the emergence of synchronized subnetworks similar to the super-paramagnetic phase in statistical physics, a paradigm that has been proven to be computationally extremely efficient, e.g., for clustering \cite{HLC1}. We expect such approaches to become pivotal for the understanding the claimed increased computational efficacy of deep layer neural networks \cite{Hintondeeplayer}, or for the explanation of the emergence of spontaneous otoacoustic emissions in the cochlea.



\bibliography{coupledHopf_database}

\end{document}